\documentclass{llncs}

\newcommand{\toolname}{FShell plugin for RVS} 

\usepackage[cmex10]{amsmath}
\usepackage{algorithmic}
\usepackage{multirow}
\usepackage{pbox}
\usepackage{listings}
\usepackage{url}
\usepackage{hyperref}
\usepackage{booktabs}
\usepackage{wrapfig}

\usepackage{paralist}

\usepackage{tikz}
\usetikzlibrary{positioning,shapes.geometric}
\usetikzlibrary{calc}
\usetikzlibrary{arrows}
\tikzstyle{parallelogram} = [trapezium, trapezium left angle=70, trapezium right angle=110,text centered, draw=black]

\begin{document}

\title{Assisted Coverage Closure\thanks{
The research leading to these results has received funding from the
ARTEMIS Joint Undertaking under grant agreement number 295311
\href{http://vetess.eu/}{``VeTeSS''}.}}

\author{Adam Nellis\inst{1},
Pascal Kesseli\inst{2},
Philippa Ryan Conmy\inst{1},
Daniel Kroening\inst{2}, 
Peter Schrammel\inst{2},
Michael Tautschnig\inst{3}}

\institute{Rapita Systems Ltd, UK \and University of Oxford, UK \and Queen Mary University of London, UK}

%\author{
%\IEEEauthorblockN{Adam Nellis\IEEEauthorrefmark{1},
%Pascal Kesseli\IEEEauthorrefmark{2},
%Daniel Kroening\IEEEauthorrefmark{2}, 
%Philippa Ryan Conmy\IEEEauthorrefmark{1},
%Peter Schrammel\IEEEauthorrefmark{2},
%Michael Tautschnig\IEEEauthorrefmark{3}
%}
%\IEEEauthorblockA{\IEEEauthorrefmark{1}Rapita Systems Ltd, UK,
%\{anellis, pconmy\}@rapitasystems.com
%}
%\IEEEauthorblockA{\IEEEauthorrefmark{2}University of Oxford,
%  Department of Computer Science, UK,
%first.lastname@cs.ox.ac.uk}
%\IEEEauthorblockA{\IEEEauthorrefmark{2}Queen Mary University of London,
%  School of Electronic Engineering and Computer Science, UK,
%mt@eecs.qmul.ac.uk}
%}
\maketitle

\begin{abstract}
The malfunction of safety-critical systems may cause damage
to people and the environment.  Software within those systems is
rigorously designed and verified according to domain specific
guidance, such as ISO26262 for automotive safety.  This paper
describes academic and industrial co-operation in tool development to
support one of the most stringent of the requirements ---
achieving full code coverage in requirements-driven testing. 

We present a verification workflow supported by a 
tool that integrates the coverage measurement tool
RapiCover with the test-vector generator FShell. The tool assists
closing the coverage gap by providing the engineer with test vectors
that help in debugging coverage-related code quality issues and
creating new test cases, as well as justifying the presence of
unreachable parts of the code in order to finally achieve full
\emph{effective} coverage according to the required criteria.

To illustrate the practical utility of the tool, we report about an
application of the tool to a case study from automotive industry.
\end{abstract}

%===============================================================================
\section{Introduction}
%===============================================================================

Software within safety-critical systems must undergo strict design and
verification procedures prior to their deployment. The recently
published ISO26262 standard~\cite{ISO26262} describes the safety life
cycle for electrical, electronic and software components in the
automotive domain. Different activities are required at different
stages of the life cycle, helping ensure that system safety
requirements are met by the implemented design. The rigor to which
these are carried out depends on the severity of consequences of
failure of the various components. Components with automotive safety
integrity level (ASIL)~D have the most stringent requirements, and
ASIL~A the least strict. One of the key required activities for
software is to demonstrate the extent to which testing has exercised
source code, also known as code coverage. This can be a challenging
and expensive task~\cite{dupuymcdc}, with much manual input required
to achieve adequate coverage results.

This paper presents work undertaken within the \textbf{Ve}rification
and \textbf{Te}sting to Support Functional \textbf{S}afety
\textbf{S}tandards (VeTeSS) project, which is developing new tools and
processes to meet ISO26262. The main contribution of this paper is an
integration of the FShell tool~\cite{HSTV08} with an industrial code
coverage tool (RapiCover) in order to generate extra test cases and
increase code coverage results. An additional contribution is to
present a discussion as to how this technology might be most
appropriately used within the safety life cycle. Achieving 100\% code
coverage can be a complex and difficult task, so tools to assist the
process are desirable, however there is a need to ensure that any
additional automatically generated tests still address system safety
requirements.

Safety standards require different depths of coverage depending on the
ASIL of the software. The requirements of ISO26262 are summarized in
Tab.~\ref{tab:ISO26262CovReqs}. The aim of requirements-based
software testing is to ensure the different types of coverage are
achieved to 100\% for each of the categories required. In practice
this can be extremely difficult, e.g.~defensive coding can be
hard to provide test vectors for. Another example is code that may be
deactivated in particular modes of operation. Sometimes there is not
an obvious cause for lack of coverage after manual review. In this
situation, generating test vectors automatically can be beneficial to
the user providing faster turnaround and improved coverage results.

%%%%%%%%%%%%%%%%%%%%%%%% ISO 26262 requirements %%%%%%%%%%%%%%%%%%%%%%%%%%
\begin{table*}[t]
	\centering
		\begin{tabular}{p{0.15\textwidth}p{0.55\textwidth}@{\quad}p{0.27\textwidth}}
      \toprule
			Type & Description & ASIL \\
      %\hline
			\midrule
			Function (arch level) & Each function in the code is exercised at least once & A, B (R); C, D (HR) \\
			%\hline
			Statement & Each statement in the code is exercised at least once & A, B (HR); C, D (R) \\
			%\hline
			  Branch &  Each branch in the code has been exercised for every outcome at least once. & A (R); B, C, D (HR) \\
        %\hline
				MC/DC & Each possible condition must be shown to independently affect a decision's outcome. & A, B, C (R); D (HR) \\
				%\hline
        \bottomrule
		\end{tabular}\vspace*{2ex}
	\caption{ISO26262 Coverage Requirements (HR = Highly Recommended, R =
  Recommended)}
	\label{tab:ISO26262CovReqs}
\end{table*}
%%%%%%%%%%%%%%%%%%%%%%%%%%%%%%%%%%%%%%%%%%%%%%%%%%%%%%%%%%%%%%%%%%%%%%%%

This paper is laid out as follows. In Sec.~\ref{sec:cc} we provide background to
the coverage problem being tackled, and criteria for success. In
Sec.~\ref{sec:impl} we describe the specific tool integration.
Sec.~\ref{sec:exp} describes an industrial automotive case study.
Sec.~\ref{sec:improv} looks at both previous work and some of the lessons
learned from the implementation experience, and suggested improvements. Finally
we present conclusions and further work. 

The contribution of this paper is by and large of practical
nature: the integration of formal-methods based tools with industrial
testing software. In the safety-critical domain these two areas are
generally separated from one another, with formal methodology used
only for small and critical sections of software to prove correctness
and viewed as an expensive procedure. In some cases the methods are
seen in direct odds to one another~\cite{Galloway}. The tool is at a
prototype stage of development, and the authors are working with
industrial partners to assess future improvements to prepare its
commercialization, as described in Sec.~\ref{sec:improv}.

%===============================================================================
\section{Assisted Coverage Closure}\label{sec:cc}
%===============================================================================

Testing has to satisfy two objectives: it has to be effective, and it
has to be cost-effective. Testing is effective if it can distinguish a
correct product from one that is incorrect. Testing is cost-effective
if it can achieve all it needs to do at the lowest cost (which usually
means the fewest tests, least amount of effort and shortest amount of
time).

Safety standards like ISO26262 and DO-178B/C demand
requirements-driven testing to increase confidence in
correct behavior of the software implemented. Correct behavior means
that the software implements the behavior specified in the
requirements \emph{and} that it does not implement any unspecified
behaviors. 
As a quality metrics they demand the measurement of \emph{coverage}
according to certain criteria as listed in
Tab.~\ref{tab:ISO26262CovReqs}, for instance.
The rationale behind using code coverage as a quality metrics for
assessing the achieved requirements coverage of a test suite is the
following:
Suppose we have a test suite that presumably covers each case in the
requirements specification, then, obviously, missing or erroneously
implemented features may be observed by failing test cases, whereas
the lack of coverage, e.g.~according to the MC/DC criterion,
indicates that there is behavior in the software which is not
exercised by a test case. This may hint at the following software and
test quality problems:
\begin{compactenum}[(A)]
\item Some cases in the requirements specification have been
  forgotten. These requirements have to be covered by additional test
  cases.
\item Features have been implemented that are not needed. Unspecified
  features are not allowed in safety-critical software and have to be
  removed.
\item The requirements specification is too vague or ambiguous to
  describe a feature completely. The specification must be
  disambiguated and refined.
\item Parts of the code are unreachable. The reasons may be:
  \begin{compactenum}[(1)]
  \item A programming error that has to be fixed. 
  \item Code generated from high-level models often contains
  unreachable code if the code generator is unable to eliminate
  infeasible conditionals. 
  \item It may actually be intended in
  case of defensive programming and error handling. 
  \end{compactenum}
  In the latter case, fault injection testing is required to exercise
  these features \cite{JH11}.
  Dependent on the policy regarding unreachable code, case (2) can be
  handled through justification of non-coverability, tuning the model or
  the code generator, or post-processing of generated code.
\end{compactenum}
The difficulty for the software developer consists in 
distinguishing above cases. This is an extremely time consuming
and, hence, expensive task that calls for tool assistance.

\subsection{Coverage Closure Problem}\label{sec:ccproblem}

\noindent Given 
\begin{compactitem}
\item an implementation under test (e.g.~C code generated from a
  Simulink model),
\item an initial test suite (crafted manually or generated by some
  other test suite generation techniques), and
\item a coverage criterion (e.g.~MC/DC),
\end{compactitem}
we aim at increasing \emph{effective} test coverage by automatically
\begin{compactitem}
\item generating test vectors that help the developer debug the software in
  order to distinguish above reasons (A)--(D) for missing coverage;
\item in particular, suggesting additional test vectors that help the developer create test cases to complete requirements coverage in case (A); 
\item proving infeasibility of non-covered code, thus giving evidence for arguing non-coverability. 
\end{compactitem}

Note that safety standards like to DO-178C~\cite{DO178BCGuide} allow
only requirements-driven test-case generation and explicitly
\emph{forbid} to achieve full structural code coverage by blindly
applying automated test-vector generation.
This can easily lead to confusion if the distinction between test-\emph{case}
generation and test-\emph{vector} generation is not clearly made.
Test-\emph{vector} generation can be applied blindly to achieve full
coverage, but it is without use by itself.  A test vector is only a
\emph{part} of a test case because it lacks the element that provides
information about the correctness of the software, i.e.~the expected
test result. 
Only the requirements can tell the test engineer what the expected
test result has to be.  Test-\emph{case} generation is thus
\emph{always} based on the requirements (or a formalized model thereof
if available). Our objective is to provide assistance for test-case
generation to bridge the coverage gap.

\subsection{Coverage Measurement}\label{sec:covmeas}

Combining a test-case generator with a coverage tool provides immediate access
to test vectors needed to obtain the level of coverage required for your
qualification level.

Coverage tools determine which parts of the code have been executed
by using instrumentation. Instrumentation points are automatically
inserted at specific points in the code. If an instrumentation point
is executed, this is recorded in its execution data. After test completion,
the coverage tool analyzes the execution data to determine 
which parts of the source code have been executed.
The tool then computes the level of coverage achieved by the tests.

We use the coverage tool RapiCover, which is part of the RVS tool suite developed by Rapita Systems Ltd.

\subsection{Test Vector Generation by Bounded Model Checking}\label{sec:bmc}

%%%%%%%%%%%%%%%%% Algorithm 
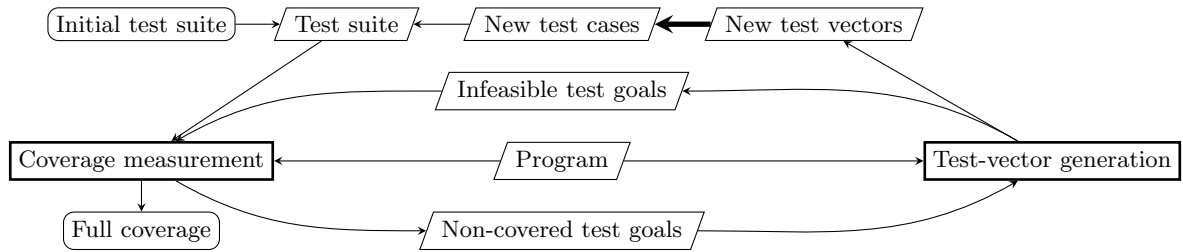
\begin{figure*}[t]
\centering
\small
\hspace*{-5em}
\begin{tikzpicture}[node distance=12em,->,>=stealth]
\matrix
{
\node[rectangle, rounded corners, draw] (start) {\small Initial test suite}; & 
\node[parallelogram, draw] (testsuite)  {\small Test suite}; &
\node[parallelogram, draw] (testcase)  {\small New test cases}; &
\node[parallelogram, draw] (testvec)  {\small New test vectors}; & \\[3ex]
&& \node[parallelogram, draw] (infeasible)  {\small Infeasible test goals}; && \\[3ex]
\node[rectangle, draw, line width=1pt] (covmeas) {\small Coverage measurement}; &&
\node[parallelogram, draw] (program)  {\small Program}; &&
\node[rectangle, draw, line width=1pt] (testgen) {\small Test-vector generation}; \\[3ex]
\node[rectangle, rounded corners, draw] (stop) {\small Full coverage}; && 
\node[parallelogram, draw] (noncov)  {\small Non-covered test goals}; && \\
 \\
};
\draw (program) to (covmeas);
\draw (program) to (testgen);
\draw (covmeas) edge[out=-30,in=180] (noncov);  
\draw (noncov) edge[out=0,in=-150] (testgen);
\draw (testgen) to (testvec);
\draw[line width=2pt] (testvec) to (testcase);
\draw (testcase) to (testsuite);
\draw (testsuite) to (covmeas);
\draw (testgen) edge[out=150,in=0] (infeasible);
\draw (infeasible)  edge[out=180,in=30] (covmeas);
\draw (start) to (testsuite);
\draw (covmeas) to (stop);
\end{tikzpicture}
\caption{\label{fig:algo}
The Coverage Closure Process
}
\end{figure*}
%%%%%%%%%%%%%%%%%

We use the test vector generator, FShell~\cite{HSTV08} (see
Sec.~\ref{sec:fshell} for details), which is based on the Software
Bounded Model Checker for C programs, CBMC~\cite{CKL04}.

Viewing a program as a transition system with initial states described
by the propositional formula $\mathit{Init}$, and the transition relation
$\mathit{Trans}$, Bounded Model Checking (BMC)~\cite{BCCZ99} can be used to
check the existence of a path $\pi$
of length $k$ from $\mathit{Init}$ to another set of states described by the formula $\psi$.
This check is performed by deciding satisfiability of the following
formula using a SAT or SMT solver:
\begin{equation}\label{equ:bmc}
\mathit{Init}(s_0)\wedge\bigwedge_{0\leq j<k} \mathit{Trans}(s_j,i_j,s_{j+1}) \wedge\psi(s_k)
%\bigvee_{0\leq j\leq k} \psi(s_k)
\end{equation}
If the solver returns the answer ``satisfiable'', it also provides a
satisfying assignment to the variables
$(s_0,i_0,s_1,i_1,\ldots,s_{k-1},i_{k-1},s_k)$.
The satisfying assignment represents one possible path $\pi=\langle
s_0,s_1,\ldots,s_k\rangle$ from $\mathit{Init}$ to $\psi$ and identifies the
corresponding input sequence $\langle i_0,\ldots,i_{k-1}\rangle$.

Besides being useful for refuting safety properties (where~$\psi$
defines the error states), BMC can be used for generating test vectors
(where $\psi$ defines the test goal to be covered).

The analysis performed by CBMC is bit-exact w.r.t.\ the machine
semantics of the execution target and CBMC provides full bit-exact
support for floating point arithmetic. Architecture-specific settings
can be configured via command line in FShell and RapiCover supports
on-target coverage measurement. We are hence guaranteed that the
generated test vectors are going to cover the test goals.
In addition, using BMC in a test-vector generator permits
generating the shortest test vectors possible to cover a
certain test goal or even a whole group of test goals, which helps
keeping test suites concise and test execution fast~\cite{SMK13}.

An advantage of using a model checker is also its ability to find test
vectors for corner cases (``Under which conditions can this floating
point variable take the value NaN?''). Moreover, in our experience,
due to the high precision of the analysis, it is even very likely to
discover inconsistencies and holes in the requirements specification
during test-vector generation.

At last, BMC can give a proof of unreachability of a test goal if
loops can be unrolled completely; or otherwise, $k$-induction~\cite{SSS00}, a
BMC-based technique for unbounded model checking, can be used to attempt a
proof.

\subsection{The Coverage Closure Process}\label{sec:ccalg}

The algorithm that we implement to assist the coverage closure process
is shown in Fig.~\ref{fig:algo}.
It proceeds as follows:
\begin{compactenum}
\item We start with an \emph{initial test suite} that has been crafted
  manually or has been generated using other test-case generation
  techniques like directed random testing. The initial test suite may
  be empty, but many test goals can be easily
  covered using test-case generation methods that are cheaper than
  Bounded Model Checking. It is thus recommended to start with such a
  base test suite.
\item In the next step, this \emph{test suite} is run using the
  \emph{coverage measurement} tool in order to obtain a list of
  \emph{non-covered test goals}. Coverage measurement can be performed
  on a developer machine to obtain approximate coverage, but final certification
  data has to be obtained by running the test suite on the actual target
  platform.
\item The \emph{test-vector generator} takes the list of non-covered
  test goals and tries to compute input values to cover them.
  Ideally, the test-vector generator is parametrized with the
  architectural parameters of the target platform in order to obtain
  guarantees that the goals are indeed going to be covered.
  As our test-vector generator is a Bounded Model Checker,
  there will be three possible outcomes of an attempt to cover
  test goals: 
\begin{compactenum}
\item A test goal has been covered. In this case this \emph{new test
  vector} is presented to the user who has to turn it into a \emph{new
  test case} to be added to the \emph{test suite}. Note that building
  the new test case is the only part of the process (bold edge) that
  is not fully automatic since human judgment is required to identify why the
  corresponding test goal has not been covered in the first place,
  i.e.~distinguishing reasons (A)--(D) in Sec.~\ref{sec:cc}.
\item It is \emph{infeasible to cover a test goal}. This happens when
  the test-vector generator comes up with a proof of
  unreachability of the test goal. As mentioned above, a Bounded Model
  Checker can provide such proofs if the loops have been unwound
  completely, for instance. In this case, the corresponding test goal
  can be annotated in the coverage report as \emph{proven infeasible}
  to justify its non-coverability. This increases \emph{effective}
  coverage by reducing the number of genuinely coverable test goals.
\item The goal has not been covered and we were unable to prove
  infeasibility of the test goal. With a Bounded Model Checker this
  can happen if the chosen bound $k$ has been too low. In this case
  the test goal will remain uncovered and it can be tried to cover it
  with a higher value for $k$ in the next iteration of the process.
\end{compactenum}
\item Coverage of the enhanced test suite is then measured again
  to identify test goals that remain uncovered, and the
  process is repeated. Generated tests typically will cover more
  test goals than intended. Measuring coverage between generating
  tests increases cost-effectiveness of the process by eliminating
  unnecessary test-case generations.
\item If there are no more non-covered test goals we have
  achieved \emph{full coverage} and the process terminates.
\end{compactenum}

Note that the process depicted in Fig.~\ref{fig:algo} is not
specific to our tool but applies in general.  
In particular, it does not rely on the test-vector generator to
guarantee that a generated test vector covers the test goal it has
been generated for, because the coverage measurement tool will check
all generated test cases anyway for
increasing the coverage.  
However, the generation of useless test cases can be avoided by 
using a tool such as FShell that can provide such guarantees.

Then, in theory, termination of the process achieving full coverage can be
guaranteed, because embedded software is finite state. In practice, however,
this depends on the reachability diameter of the system~\cite{KS03} and the
capacity of the test-vector generator to cope with the system's size and complexity.

%===============================================================================
\section{\toolname{} Implementation}\label{sec:impl}
%===============================================================================

The input to the tool%
\footnote{RVS is licensed software. An evaluation version can be requested from \href{http://www.rapitasystems.com}{\url{http://www.rapitasystems.com}}. 
The licensing policy disallows anonymous licenses. To compensate for this, we provide a video showing the plug-in here: \href{https://drive.google.com/file/d/0B7xeLJ8vk3W8Y094TVc4Rmh0S0k}{\url{https://drive.google.com/file/d/0B7xeLJ8vk3W8Y094TVc4Rmh0S0k}}.}
 is a C program with an initial test suite.  The output of the tool is
 twofold.  The first output is a set of generated test vectors that
 augment the initial test suite to increase its coverage.  The second
 output is a coverage report detailing the level of coverage achieved
 by the initial test suite, and the extra coverage added by the
 generated test cases.

\setlength{\intextsep}{5pt}%
\begin{wrapfigure}{r}{0.66\textwidth}
%\begin{figure}[t]
	\centering
		\includegraphics[width=8cm]{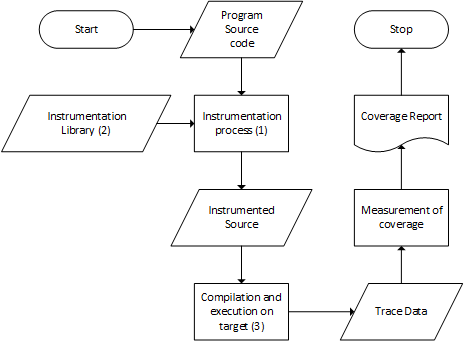}
	\caption{RVS Process}
	\label{fig:RVSProcess}
%\end{figure}
\end{wrapfigure}

\begin{figure*}
  \centering
\includegraphics[width=0.86\textwidth]{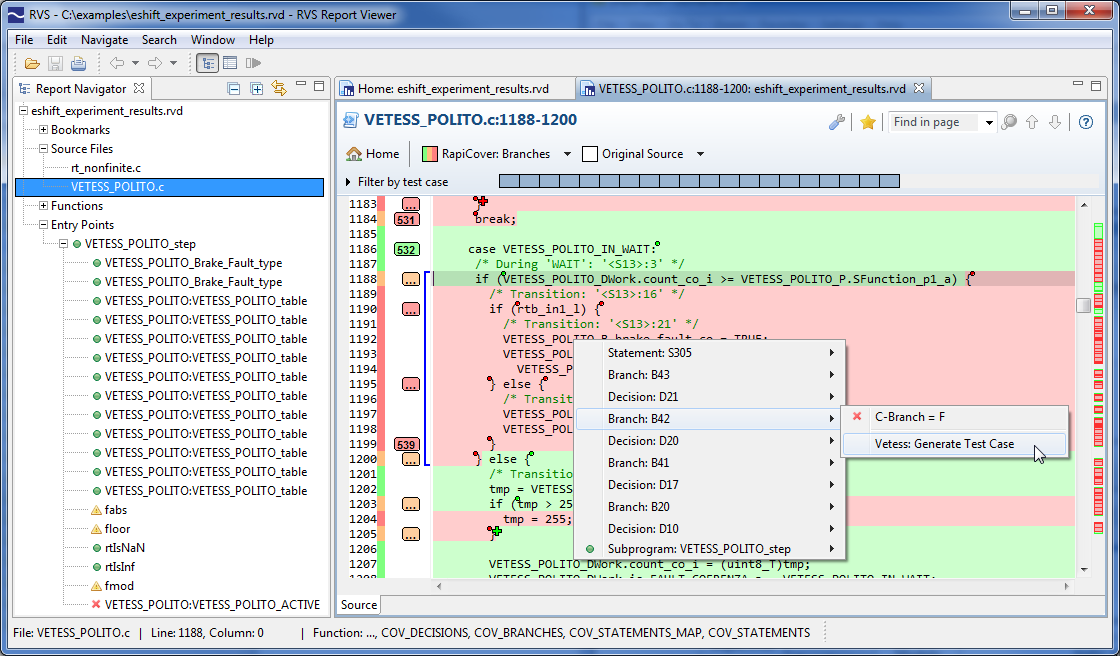}
\caption{\label{fig:plugin}
Screenshot of RapiCover with the FShell Plug-in
}
\end{figure*}

FShell has been integrated into RapiCover as context menu option,
shown in Fig.~\ref{fig:plugin}.  RapiCover can be used to select a
single function, call, statement, decision or branch.  The tool then
uses FShell to generate a test vector for this element.
Alternatively, the tool has a button to generate as much coverage as
possible.  When this option is chosen, the tool goes around the loop
described in Fig.~\ref{fig:algo}, using FShell to repeatedly generate
test cases to increase the coverage as much as possible, verifying the
obtained coverage with RapiCover.

There is tension between the need to demonstrate that the activities
prescribed by ISO26262 have been met in spirit as well as with
quantifiable criteria. Recall that achieving 100\% code coverage
during testing does not ensure the code meets its intent. Consequently
the FShell plug-in would be provided as advisory
service, generating candidate test vectors, which a user can examine to
help them identify why their planned testing was inadequate. Values generated
need to be assessed for being valid for
the system under test, i.e.~reflect real world values
that could be input to a function, e.g.~from a sensor.

%%%%%%%%%%%%%%%%% Plugin 
%%%%%%%%%%%%%%%%%

%Availability... refer to \url{http:\\...} for evaluation
%setup... \pscmt{should be made available to reviewers, if possible}

\subsection{Introduction to RapiCover}

RapiCover%
\footnote{\url{http://www.rapitasystems.com/products/rapicover}}
 uses instrumentation to determine which program parts have been
executed. Instrumentation points are automatically inserted at specific points
in the code. Execution of an instrumentation point is recorded in
its execution data. Upon test completion, RapiCover analyzes the execution data
to determine which instrumentation points have been hit.

The first step in the RapiCover analysis process is to create an
instrumented build of the application ((1) in
Fig.~\ref{fig:RVSProcess}).  RapiCover automatically adds
instrumentation points ((2) in Fig.~\ref{fig:RVSProcess}) to the
source code.

The instrumentation code itself takes the form of very lightweight
measurement code that is written for each target to ensure minimal
impact on the performance of the software, and to support on target
testing for environments with limited resources.
The instrumented software and possibly an instrumentation library are
compiled and linked using the standard compiler tool chain. The
executable produced is then downloaded onto the target hardware. The
executable is exercised and instrumentation data ((3) in
Fig.~\ref{fig:RVSProcess}) is generated and retrieved. This data is
used to generate coverage metrics.

\subsection{Introduction to FShell}\label{sec:fshell}

FShell%
\footnote{Available from: \url{http://forsyte.at/software/fshell}} 
is an extended testing environment for C programs supporting a rich
scripting language interface. FShell's interface is designed as a
database engine, dispatching queries about the program to various
program analysis tools.  These queries are expressed in the FShell
Query Language (FQL).
Users formulate test specifications and coverage criteria, challenging
FShell to produce test suites and input assignments covering the
requested patterns. The program supports a rich and extensive
interface. The expressions used for the \toolname{} implementation are
listed in Tab.~\ref{tab:fshell-syntax} with syntax and examples.

\setlength{\intextsep}{5pt}%
\begin{wraptable}{r}{0.72\textwidth}
%\begin{table}[t]
\small
\centering
\begin{tabular}{ l c l }
  \toprule
Expression Name & Syntax & Example\\
  \midrule
Function Call & @CALL(\ldots) & @CALL(X)\\
Concatenation & . & @CALL(X).@CALL(Y)\\
Sequence & -\textgreater & @CALL(X)-\textgreater@CALL(Y)\\
Negation & ``NOT(\ldots)'' & ``NOT(@CALL(X))''\\
Repetition & * & @CALL(X)*\\
Alternative & + & (@CALL(X) + @CALL(Y)) \\
  \bottomrule
\end{tabular}
\vspace*{0.5em}
\caption{\label{tab:fshell-syntax}
FShell expressions}
%\end{table}
\end{wraptable}

\emph{@CALL(X)} requires generated test cases to call
function \emph{X}. This is the only primitive expression used in the module. The
concatenation operator \emph{.} joins two expressions, requiring them to
be satisfied subsequently. As an example, a test case
generated by \emph{@CALL(X).@CALL(Y)} covers a call to \emph{X} immediately
followed by \emph{Y}. This is similar to the sequence operator
\emph{-\textgreater}, which requires the second call to occur
eventually. \emph{@CALL(X)-\textgreater@CALL(Y)} is thus fulfilled if a call to
\emph{X} is eventually followed by a call to \emph{Y}. The negation
\emph{``NOT(@CALL(X))''}
is satisfied by every statement except a call to function \emph{X}. The
repetition operator is implemented along the lines of its regular expression
pendant, such that \emph{@CALL(X)*} is satisfied by a series of calls to
\emph{X}. Finally, the alternative operator implements logical disjunction, such
that \emph{(@CALL(X) + @CALL(Y))} will be satisfied if either a call to \emph{X}
or \emph{Y} occurs.

The expressions and operators above are all that is used by the FShell plug-in
to generate the test vectors requested by RapiCover. Sec.~\ref{sec:rvsfshell}
illustrates how these expressions are used to convert test goals to equivalent
FQL queries.

\subsection{Use of FShell within RapiCover}\label{sec:rvsfshell}

%%%%%%%%%%%%%%%%% FShell Java module architecture
\setlength{\intextsep}{5pt}%
\begin{wrapfigure}{r}{0.6\textwidth}
%\begin{figure}[t]
\small
\centering
\begin{tikzpicture}[node distance=2.2em and 6em,>=stealth,
every node/.style={minimum size=2.2em}]
\node[rectangle, draw] (java) at (0,0)
{\parbox{5em}{\centering \small Java \\
Module}}; \node[rectangle, draw, below=of java] (fshell)
{\parbox{5em}{\centering \small FShell}}; 
\node[rectangle, draw, left= of java] (rapi) 
{\parbox{5em}{\centering \small RapiCover}};  
\draw[<-] ([yshift=-0.5em]rapi.east) to ([yshift=-0.5em]java.west)
node [above left,xshift=-0.7em,yshift=0.4em] {\small Test Goals};
\draw[<-] ([xshift=-0.5em]java.south) to ([xshift=-0.5em]fshell.north)
node [above right,xshift=1em,yshift=0.0em] {\small FQL Queries};
\draw[->] ([yshift=0.5em]rapi.east) to ([yshift=0.5em]java.west)
node [below left,xshift=-0.0em,yshift=-0.4em] {\small Test Vectors};
\draw[->] ([xshift=0.5em]java.south) to ([xshift=0.5em]fshell.north)
node [above left,xshift=-1em,yshift=0.0em] {\small Test Suites};
\end{tikzpicture}
\caption{\label{fig:javamodule}
Architecture of \toolname{}
}
%\end{figure}
\end{wrapfigure}
%%%%%%%%%%%%%%%%%

The \toolname{} translates test goals requested by RapiCover into
FQL queries covering these goals in FShell, as
illustrated in Fig.~\ref{fig:javamodule}. Test goals are specified using
marker elements from the RapiCover instrumentation, which can identify
arbitrary statements in the source code by assigning them an
\emph{instrumentation point id}. In accordance with MC/DC criteria,
decisions and their constituting conditions are further identified using unique
\emph{decision and condition point ids}.

\begin{figure}[t]
\begin{minipage}{0.5 \textwidth}
\footnotesize
\begin{lstlisting}[language=c]
int main() {
    // ...
    if(a == b || b != c) {
        printf("%d %d\n", a, b);
    }
    return 0;
}
\end{lstlisting}
\vspace{4.2em}
\end{minipage} 
\begin{minipage}{0.5 \textwidth}
\footnotesize
\begin{lstlisting}[language=c]
int main() {
    // ...
    Ipoint(1);
    if(Ipoint(4, Ipoint(2, a == b) ||
       Ipoint(3, b != c))) {
      Ipoint(5);
      printf("%d %d\n", a, b);
    }
    Ipoint(6);
    return 0;
}
\end{lstlisting}
\end{minipage}
\caption{\label{lst:instrumented-code-example}
Code example before and after after RapiCover instrumentation}
\end{figure}

Fig.~\ref{lst:instrumented-code-example} shows an example program before and
after RapiCover instrumentation. The module supports two categories of test
goals: \emph{Instrumentation Point Path Test Goals} and \emph{Condition Test
Goals}. The former specifies a simple series of instrumentation points to be
covered by FShell. The system also permits \emph{inclusive or} and
\emph{negation} operators in instrumentation point paths, allowing to specify a
choice of instrumentation points to be covered or to make sure that a requested
instrumentation point is not covered by the provided test vector. As an example,
the instrumentation point path \emph{1-\textgreater5-\textgreater6} in
Fig.~\ref{lst:instrumented-code-example} is only covered if the decision in
the \emph{if} statement evaluates to \emph{true}.
Conversely, the path \emph{1-\textgreater NOT(5)-\textgreater6} is only covered
if it evaluates to \emph{false}. The former can be achieved with inputs $a{=}1,
b{=}1, c{=}2$, whereas the latter could be covered using the input
vector $a{=}1, b{=}2, c{=}2$.
\emph{Condition Test Goals} on the other hand are specified by a single
\emph{decision point} and multiple \emph{condition points}, as well as the
desired truth value for each decision and condition. This allows us to cover
branch conditions with precise values for its sub-conditions. As an
example, the condition test goal
\emph{(4,true) \mbox{-\textgreater} (2,false) \mbox{-\textgreater} (3,true)} would be covered by
the input vector $a{=}1, b{=}2, c{=}3$.

\begin{table}[b!]
\small
\centering
\begin{tabular}{ p{8em} p{6em} p{22em} }
  \toprule
  Category & Goal & FQL \\\midrule
  \multirow{3}{*}{\pbox{8em}{
  \hyphenation{In-stru-men-ta-tion} Instrumentation\\ Point Path Goal}} &
  Simple &
  \pbox{21em}{\vspace{0.2em}
  @CALL(Ipoint5) -\textgreater
  @CALL(Ipoint6) -\textgreater
  \ldots\vspace{0.3em}}\\\cmidrule{2-3}
   & Disjunction &
   \pbox{21em}{\vspace{0.2em}
   ( @CALL(Ipoint5)
   + @CALL(Ipoint6)
   + \ldots)\vspace{0.3em}}\\\cline{2-3}
   & Complement &
  \pbox{21em}{\vspace{0.2em}
  @CALL(Ipoint1)."NOT(@CALL(Ipoint5))*".\\
  @CALL(Ipoint6)-\textgreater
  \ldots\vspace{0.3em}}\\\midrule
  \multirow{2}{*}{\pbox{8em}{Condition Goal}}
   & Condition &
  \pbox{21em}{\vspace{0.2em}
   @CALL(Ipoint2f)."NOT(@CALL(Ipoint1))*".\\
   @CALL(Ipoint2t)."NOT(@CALL(Ipoint1))*".\\
   +\ldots \vspace{0.3em}}
\\\cmidrule{2-3}
   & Decision &
   \pbox{21em}{\vspace{0.2em}
   @CALL(Ipoint4t)}
\\\bottomrule
\end{tabular}
\vspace*{0.5em}
\caption{\label{tab:goals-to-queries}
Test Goal Types and FShell Queries}
\end{table}

The instrumentation elements introduced by RapiCover need to be mapped to an
equivalent FQL query using the features presented in Tab.~\ref{tab:fshell-syntax}. For this purpose, we replace their default
implementation in RapiCover by synthesized substitutions which are optimized for
efficient tracking by FShell. These mock implementations are synthesized for
each query and injected into the program on-the-fly at analysis time. Standard
FQL queries are then enough to examine these augmented models for the specified
coverage goals. Tab.~\ref{tab:goals-to-queries} shows explicitly how these goals
can described using the FShell query syntax.

%===============================================================================
\section{Evaluation}\label{sec:exp}
%===============================================================================
The \toolname{} has been tested using an industrial automotive use case,
for a software managed controller.

\subsection{Case Study: e-Shift Park Control Unit}\label{sec:casestudy}

To illustrate the features and utility of the tool,
we applied it to the software of an e-Shift Park Control Unit.
This system%
\footnote{The C code was provided by Centro Ricerche Fiat under a
  GPL-like license and can be downloaded here:
  \href{https://drive.google.com/file/d/0B22MA57MHHBKamhQMmpEQlRWVG8}{\url{https://drive.google.com/file/d/0B22MA57MHHBKamhQMmpEQlRWVG8}}. The
  C code was generated from a Simulink model that has not been
  disclosed, unfortunately.}
is in charge of the management of the mechanical park lock that
blocks or unblocks the transmission to avoid unwanted movement of the vehicle
when stopped.  The park mode is enabled either by command of the driver via
the gear lever (PRND: park/rear/neutral/drive) or automatically.

%%%%%%%%%%%%%%%%% Case Study Illustration 
\setlength{\intextsep}{5pt}%
\begin{wrapfigure}{r}{0.7\textwidth}
%\begin{figure}[t]
\small
\centering
\begin{tikzpicture}[node distance=2em and 3em,>=stealth]
\node[rectangle, draw] (vcu) at (0,0) 
{\parbox{5em}{\centering \small vehicle \\ control \\ unit}}; 
\node[rectangle, draw, left=of vcu] (dashboard) 
{\parbox{5em}{\centering \small dashboard}}; 
\node[rectangle, draw, right= of vcu, fill=gray] (park) 
{\parbox{5em}{\centering \small e-Park \\ Control \\ Unit}}; 
\node[rectangle, draw, below= of dashboard] (prnd) 
{\parbox{5em}{\centering \small PRND\\ switches}}; 
\node[rectangle, draw, below= of vcu] (powertrain)  {\small powertrain}; 
\node[rectangle, draw, below= of park] (parklock)  {\small park lock}; 
\draw[<-] (dashboard) to (vcu);
\draw[->] (prnd) to (vcu);
\draw[<->] (powertrain) to (vcu);  
\draw[<->] (vcu) to (park);
\draw[<-] (vcu) to (parklock);
\draw[<->] (park) to (parklock);
\end{tikzpicture}
\caption{\label{fig:casestudy}
Case Study: e-Shift Park Control Unit
}
%\end{figure}
\end{wrapfigure}
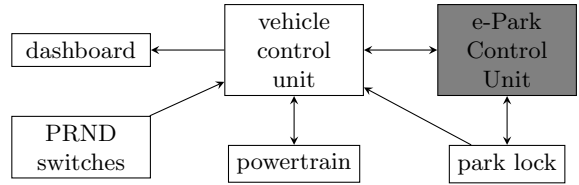
%%%%%%%%%%%%%%%%%

Fig.~\ref{fig:casestudy} shows the architectural elements the e-Park
system is communicating with.  The vehicle control unit monitors the
status of the vehicle via sensors and informs the driver, in
particular, about the speed of the vehicle and the status of the gears
via the dashboard.  The e-Park Control Unit is responsible for taking
control decisions when to actuate the mechanical park lock system.

Among many others, the following requirements have to be fulfilled:
\begin{compactenum}
\item Parking mode is engaged if vehicle speed is below 6 km/h and the driver
  presses parking button (P) and brake pedal.
\item If vehicle speed is above 6 km/h and the driver presses the parking button
  (P) and brake pedal then commands from the accelerator pedal are ignored;
  parking mode is activated as soon as speed decreases below 6 km/h.
\item If vehicle speed is below 6 km/h and the driver presses the
  driving button (D) and brake pedal, then forward driving
  mode is enabled.
\item If vehicle speed is above 6 km/h then backward driving
  mode (R) is inhibited.
\end{compactenum}

As is typical for embedded software, the e-Park Control Unit software
consists of tasks that --- after initialization of the system on
start-up --- execute periodically in the control loop until system
shutdown.
A test vector hence consists of a sequence of input values
(sensor values and messages received via the communication system)
that may change in each control loop iteration. We call the 
number of iterations the \emph{length} of the test vector.

To generate valid test vectors, a model of the
vehicle is required. Otherwise, the test vector generator may produce
results that are known not to occur in the running system, such as
infinite vehicle velocity.
For the case study this model consisted of assumptions about the input
value ranges, such as ``The speed of the car will not exceed
1000 km/h, or reduce below 0 km/h.'' These assumptions are part of the
admissible operating conditions as stated in the requirements
specification.

\subsection{Experimental Setup}\label{sec:expdesign}

%\begin{figure*}[t]
%  \centering
%\includegraphics[width=0.86\textwidth]{coverage_before.png}
%\caption{\label{fig:results1}
%Experimental results: Coverage of the initial test suite
%}
%\end{figure*}
 
\begin{table}[b!]
\begin{tabular}{@{}r@{\;}|@{\;}p{0.45\textwidth}@{\;}|@{\;}p{0.45\textwidth}}
   & \toolname{} & random search + reduction \\
\hline
1. & \multicolumn{2}{p{0.9\textwidth}}{\centering Start with the initial test suite.} \\
2. & \multicolumn{2}{p{0.9\textwidth}}{\centering Compile and run the C source code with the current test suite, using RapiCover to generate a coverage report.} \\
3. & RapiCover provides FShell with a list of non-covered test goals. & \\
4. & FShell generates a test vector for these non-covered test goals. & Generate  a random test vector, uniformly distributed over the admissible input
ranges. \\
5. & FShell feeds back information about infeasible test goals and test 
vectors for feasible test goals. & \\
6. & \multicolumn{2}{p{0.9\textwidth}}{\centering Create C test cases based on these test vectors. } \\
7. & \multicolumn{2}{p{0.9\textwidth}}{\centering Re-compile and re-run the C code
     with this new test case, using RapiCover to verify that the
     generated test case does indeed cover the test goal.} \\
8. &  & If the coverage has increased then keep the test case; otherwise discard it. \\
9. & \multicolumn{2}{p{0.9\textwidth}}{\centering Repeat from step 3.} 
\end{tabular}~\\
\caption{\label{tab:setup}
Experimental setup of the two approaches that we compare.}
\end{table}

In order to evaluate the \toolname{}, we used the C
source code of the e-Shift case study (approx.\ 4KLOC) and
started out with an initial test suite consisting of 100
random test vectors uniformly distributed over the admissible input
ranges. 
Then we incrementally extended this test suite by additional test vectors
generated by the following two approaches:
\begin{compactenum}
\item \toolname{} following the process illustrated in
Fig.~\ref{fig:algo}.
\item A combination of test vector generation based on random search
  and greedy test suite reduction.
\end{compactenum}
We compared the achieved coverage gain and resulting test suite sizes after running both approaches for 8~hours.
Tab.~\ref{tab:setup} describes our experimental setup.

The runtime of FShell is exponential in the loop bound of this main
loop. Choosing a too high loop bound results in FShell taking
prohibitively long to run, yet setting the loop bound too low results
in some branches not being coverable.  As mitigation, we started the
experiment with a loop bound of 1, then we gradually increased the loop bound
to cover those branches that we were not able cover in previous iterations.
As explained in Section~\ref{sec:ccproblem}, step 6 in
Tab.~\ref{tab:setup} is not automatic since it needs information from
the requirements specification. For the sake of our comparison that does
not care about the pass/fail status of the test, we skipped the manual
addition of the expected test outcome.

\subsection{Results}\label{sec:results}

\begin{table}[t]
\small
\centering
\begin{tabular}{ l || c || c | c | c | c || c }
 & Initial & \multicolumn{4}{c ||}{Random search} & FShell\\
 & test suite & \multicolumn{4}{c ||}{} & plug-in\\\hline
Runtime (hh:mm) & - & 00:33 & 01:04 & 06:15 & 08:00 & 08:00\\
Generated test cases & - & 500 & 1000 & 5000 & 6092 & 7\\
\hspace{1em}Thereof non-redundant & - & 7 & 10 & 13 & 13 & 7\\
Total test cases & \textbf{100} & 107 & 110 & 113 & \textbf{113} & \textbf{107}\\\hline
Statement coverage & \textbf{51.9\%} & 52.0\% & 52.1\% & 52.1\% & \textbf{52.1\%} & \textbf{52.5\%}\\
\hspace{1em} Increase &  & \multicolumn{1}{r|}{0.1\%} & \multicolumn{1}{r|}{0.2\%} & \multicolumn{1}{r|}{0.2\%} & \multicolumn{1}{r||}{\textbf{0.2\%}} & \multicolumn{1}{r}{\textbf{0.6\%}}\\\hline
MC/DC coverage & \textbf{28.4\%} & 30.5\% & 31.1\% & 31.9\% & \textbf{31.9\%} & \textbf{33.6\%}\\
\hspace{1em} Increase &  & \multicolumn{1}{r|}{2.1\%} & \multicolumn{1}{r|}{2.7\%} & \multicolumn{1}{r|}{3.5\%} & \multicolumn{1}{r||}{\textbf{3.5\%}} & \multicolumn{1}{r}{\textbf{5.2\%}}\\
\end{tabular}
\vspace*{0.5em}
\caption{\label{tab:results1}
Evaluation results: Comparing \toolname{} against test vectors generated by random search.}
\end{table}

We ran the experiment for 8 hours.
The first approach spent more than 99\% of this time
within FShell.  Within this time frame, the loop bound reached~2, and
thus not all branches could be covered.  Nevertheless an increased
coverage was achieved as detailed in
Tab.~\ref{tab:results1},
%and~ \ref{fig:results2} 
which shows the baseline coverage on the initial test suite (second
column from the left) and the increase in percentage of coverage
gained by the tool in this experiment (rightmost column).
The code under test implements a state machine, which is mostly
decisions with very few functions and calls, which is why we focussed
on decision and statement coverage for our evaluation.

To underpin the benefit of our tool we compared these results to the
second approach described in Tab.~\ref{tab:setup}, a random search
test generation strategy. Tab.~\ref{tab:results1} shows four snapshots
of this search (middle columns) after exploring 500, 1000, 5000 and
eventually 6092 test vectors of length 5 out of the admissible input
range.%
\footnote{We chose length 5 because it seems a good compromise between
  increasing coverage and keeping test execution times short for this
  case study: adding 100 test vectors of length 5 increased coverage
  by 1.1\%; 100 test vectors of length 10 increased it by only 1.3\%
  while test execution times would double and only half as many test
  vectors could be explored.}
 The results show that more than 99.99\% of the generated test
vectors added by the random search are redundant and do not increase
the coverage of the suite. This confirms that the system under test
represents a particularly challenging case for black-box test case
generation and that only very few test vectors in the input range lead
to actual coverage increase.

On the other hand, the \toolname{}
achieves a significantly larger increase in the same amount of time in
both statement and MC/DC coverage. In addition to this, the plug-in
achieves this increased coverage with only half as many new test
vectors as the random approach, leading to an overall smaller and more
efficient test suite.
% To underpin the benefit of the approach, we compare these results to
% the addition of 100 additional test vectors of length 5 
% through random test vector generation, which increased
% the MC/DC coverage by 1.1\%, 100 additional test vectors of length 10
% (+1.3\%), 100 of length 10 plus another 100 of length 20 (+1.9\%).
%
% This shows the high value of the 7 very short test vectors of length 1
% and 2 generated by FShell in comparison of a large number of long
% random generated ones.

This evaluation thus underlines the benefit from our tool integration 
to support the coverage closure process on an industrial case study.
The expected reduction in manual work needs to be investigated in a broader
industrial evaluation involving verification engineers performing 
the entire coverage closure process.

\section{Background Context and Applicability} \label{sec:improv}

\subsection{Novelty of the Approach}
There is much work existing for test case generation 
using Model Checking techniques~\cite{FWA09}, but a smaller amount
targeted directly at the high criticality safety domain where the criterion and
frameworks for test case generation are restricted. A useful survey relating to
MC/DC can be found in~\cite{zamli2013test}. In~\cite{GhaniMCDC} Ghani and Clark
present a search based approach to generating test frameworks. There are two
issues with the approach presented, firstly that it is applied to Java---a
language which is rarely used for safety-critical software, and particularly not
for the most critical software. The second is more subtle: the test cases were
generated to ensure that the minimal set of truth tables for MC/DC were
exercized, but without consideration of the validity of any of the test data.
Additionally, we emphasize that our approach takes into account existing coverage that
has already been achieved and complements the requirements based testing, rather
than completely replacing it. 
% A third problem is that the Ghani approach
% provides non-deterministic output, i.e.~the test cases generated are not the
% same each time it is run. Repeatability is standard criterion during safety
% critical testing.

Other work such as~\cite{Kandl} looks at modification of the original source
through mutation testing in order to assess effectiveness of the tests. This
could be considered an adjunct to our methodology, but at present mutation
testing is not widely adopted by industry. Jones~\cite{Jones} considers test
prioritization and test suite reduction, but not new test case generation.

\subsection{Wider Issues and Lessons Learnt}
In order to encourage wider adoption of this integrated tool, we need
to consider where it would fit in users' workflow and verification
processes, as well as meeting the practical requirements of the
standard. As noted earlier, fully automated code coverage testing is
not desirable as it misses the intent of the requirements based
testing process. However, achieving full code coverage is a difficult
task, and often requires a large amount of manual inspection of
coverage results to examine what was missing. Hence providing the user
with suggested test data is potentially very valuable and could
improve productivity in one of the most time consuming and expensive
parts of the safety certification process.

Another benefit of integrating test case generation and coverage
measurement is test suite reduction. The coverage measurement tool
returns for each test case a list of covered goals. Test suite
reduction is hence the computation of a minimal set cover (an
$NP$-complete problem). Approximate algorithms~\cite{TG05} may be used
to achieve this in reasonable runtimes.

FShell uses a class of semantically exact, but computationally
expensive, $NP$-complete algorithms relying on SAT solvers.  Depending
on the programs or problems posed to the solver the analysis may take
long time to complete. 
Initial feedback on the tool showed that the concept was very well
received by Automotive engineers.
Speed was considered an issue, % and should be improved in future,
however, keeping in mind that today's practice for full coverage
testing may take several person months with an estimated cost of \$100
per LOC,%
\footnote{Atego. ``ARINC 653 \& Virtualization Solutions Architectures and
Partitioning'', Safety-Critical Tools Seminar, April 2012.}
 there is great potential for cutting down time and cost spent
in verification by running an automated tool in the background for a
couple of days.

% Test data must be meaningful in the context of the system under
% test. For example, adding a test vector which reflects valid ranges from an
% input sensor. One example from our case study is ensuring the test vector
% generated provides a value for the vehicle speed which reflects real life
% performance of the car, and is within a valid range. 

Initially, we sometimes failed to validate that a test vector that was
generated to cover a test goal actually covers that test goal.  E.g.,
one reason were imprecise number representations in the test vector
output. Using the exact hexadecimal representation for floating point
constants instead of the imprecise decimal one fixed the problem.
This highlights the value of bit-exact analysis as well
as the importance of re-validating coverage using RapiCover in the
process (Fig.~\ref{fig:algo}).

Note also that this process itself is independent of the tools used
which offers a high degree of flexibility.  On the one hand, it is
planned that in future RVS will support alternative backends in place
of FShell.
On the other hand, FShell can be combined -- without changing the
picture in Fig.~\ref{fig:algo} -- with a mutation testing tool (in
place of RapiCover) to generate test vectors to improve mutation
coverage.

%===============================================================================
\section{Conclusion}\label{sec:concl}
%===============================================================================
This paper has demonstrated the successful integration of the FShell
tool with an industrial code coverage tool. Using the integrated tools
we were able to increase MC/DC code coverage of an existing, sizeable
test suite for an industrial automotive case study from 28.4\% to
33.6\%. When compared to a random black-box test vector generation
strategy, our approach was able to generate a 1.5 times higher
coverage increase within the same amount of time. Our tool achieves
this coverage gain with half as many test vectors, and these test
vectors are much shorter than those generated by random search,
leading to more more compact test suites and faster test execution
cycles. Moreover, the integration of the two tools simplifies test
case generation and coverage measurement work flows into a unified
process.

Future work will consider better integration with the debugging
environment to inspect test vectors, and warning the user about
potentially unrealistic environment assumptions such as $\infty$ for
vehicle speed. In addition, better support should be provided for
exporting the test vectors into the users' existing test suite and
testing framework.

%     Transfer of academic approaches into open source or commercial tools and sharing lessons learned from this transfer;
%     Creation of an innovative tool;
%     Description of a significant adoption effort of a verification approach with detailed descriptions of the strategies and challenges of this adoption effort;
%     Description of a working prototype of your research technique, any feedback obtained from practitioners and description of the technical challenges in developing it.

%\paragraph{Future work}
%will consider....

%\pscmt{integration with debugger to inspect test vectors}
%\pscmt{better support for exporting test vectors to existing test suite}

\bibliographystyle{splncs03}
\bibliography{biblio}

\end{document}